\def\year{2020}\relax
\documentclass[letterpaper]{article} 
\usepackage{aaai20}  
\usepackage{times}  
\usepackage{helvet} 
\usepackage{courier}  
\usepackage[hyphens]{url}  
\usepackage{graphicx} 
\usepackage{graphicx}  
\usepackage{amsmath,amsthm,amsfonts,amssymb}
\usepackage{xspace}
\usepackage[dvipsnames]{xcolor}
\usepackage{natbib}

\usepackage{algorithm} 
\usepackage{algorithmic}

\newcommand{\omt}[1]{}

\theoremstyle{definition}

\newcommand{\squishlist}{
   \begin{list}{{{\small{$\bullet$}}}}
    { \setlength{\itemsep}{3pt}      \setlength{\parsep}{1pt}
      \setlength{\topsep}{1pt}       \setlength{\partopsep}{0pt}
     \setlength{\leftmargin}{1em} \setlength{\labelwidth}{1em}
      \setlength{\labelsep}{0.5em} } }
\newcommand{\squishend}{  \end{list}  }

\def\hmath$#1${\texorpdfstring{{\rmfamily\textit{#1}}}{#1}}

\ifodd 1
\newcommand{\yl}[1]{{\color{blue}(YL: #1)}}
\else
\newcommand{\yl}[1]{}
\fi
\newcommand{\ignore}[1]{}

\title{Replication Markets: Results, Lessons, Challenges and Opportunities in AI Replication}

\author{\\
\textbf{Yang Liu$^1$, Michael Gordon$^2$, Juntao Wang$^3$, Michael Bishop$^4$, Yiling Chen$^3$,}\\ \textbf{Thomas Pfeiffer$^2$, Charles Twardy$^4$, and Domenico Viganola$^5$}
~\\
~\\
$^1$ Corresponding author: yangliu@ucsc.edu, University of California, Santa Cruz, CA USA\\
$^2$ Massey University, Auckland, New Zealand\\
$^3$ Harvard University, Cambridge, MA USA \\
$^4$ KeyW Research Lab\\
$^5$ George Mason University, Fairfax, VA USA 
}

\begin{document}
\maketitle

\begin{abstract}

The last decade saw the emergence of systematic large-scale replication projects in the social and behavioral sciences, \citep{camerer2016evaluating,camerer2018evaluating,ebersole2016many,klein2014theory,klein2018many,open2015estimating}. These projects were driven by theoretical and conceptual concerns about a high fraction of ‘false positives’ in the scientific publications \citep{ioannidis2005most}( and a high prevalence of ‘questionable research practices’ \citep{simmons2011false}. Concerns about the credibility of research findings are not unique to the behavioral and social sciences; within Computer Science, Artificial Intelligence (AI) and Machine Learning (ML) are areas of particular concern \citep{lucic2018gans,Freire:2012:CRS:2213836.2213908,gundersen2018state,henderson2018deep}. Given the pioneering role of the behavioral and social sciences in the promotion of novel methodologies to improve the credibility of research, it is a promising approach to analyze the lessons learned from this field and adjust strategies for Computer Science, AI and ML 
In this paper, we review approaches used in the behavioral and social sciences and in the DARPA SCORE project. We particularly focus on the role of human forecasting of replication outcomes, and how forecasting can leverage the information gained from relatively labor and resource-intensive replications. We will discuss opportunities and challenges of using these approaches to monitor and improve the credibility of research areas in Computer Science, AI and ML. 
\end{abstract}

\section{Introduction}
Unreproducible scientific results will potentially mislead the progress of science, and undermine the trustworthiness of the research community.  The last decade saw the emergence of systematic large-scale replication projects in the social and behavioral sciences, and other research fields (Camerer et al., 2016, 2018; Ebersole et al., 2016; Klein et al., 2014, 2018; Open Science Collaboration, 2015). These projects were driven by theoretical and conceptual concerns about a high fraction of `false positives' in the scientific publications (Ioannidis, 2005) and a high prevalence of ‘questionable research practices’ (Simmons, Nelson, \& Simonsohn, 2011). In these projects, a well-defined subset of the literature in a research field is systematically sampled, and all studies that meet pre-defined criteria are selected for replication. Replication includes a new round of data collection based on protocols that are as similar as possible to the original protocols, and an analysis of the new data that follows the approaches of the original studies.

In the social and behavioral sciences, the rates of successful replication ranged from 39\% to 62\% \citep{camerer2016evaluating,camerer2018evaluating,open2015estimating}. Successful replication is typically defined as a statistically significant effect that is in the same direction as the original effect, though most replication studies provide a number of additional characterizations, including replication effect sizes. The relatively low rates of replication have raised concerns about the credibility of research in the social and behavioral sciences which have been captured by the term “replication crisis” (Baker, 2016). 

Concerns about the credibility of research findings are not unique to the behavioral and social sciences; within Computer Science, Artificial Intelligence (AI) and Machine Learning (ML) are areas of particular concern. Unreproducible results have been reported across different sub-areas such as reinforcement learning and deep learning \citep{lucic2018gans,Freire:2012:CRS:2213836.2213908,gundersen2018state,henderson2018deep}. The recent National Academies' Report \citep{NAP25303} highlights computational reproducibility:
\emph{
\begin{quote}
    Notably, however, a number of systematic efforts to reproduce computational results across a variety of fields have failed in more than one-half of the attempts made, mainly due to insufficient detail on digital artifacts, such as data, code, and computational workflow.
\end{quote}
}
The sheer scale of some efforts like supercomputer runs on Titan mean the reproducibility is largely theoretical. We could extend this to AlphaGo, or top Starcraft systems, or to a lesser extent, GPT-2.
The report notes,
\emph
{
\begin{quote}
    Artificial intelligence and machine learning present unique new challenges to computational reproducibility, and as these fields continue to grow, the techniques and approaches for documenting and capturing the relevant parameters to enable reproducibility and confirmation of study results needs to keep pace.
\end{quote}
}
And generally,
\emph{
\begin{quote}
    The committee’s definition of reproducibility is focused on computation because of its major and increasing role in science. ... but this [data] revolution is not yet uniformly reflected in how scientists develop and use software and how scientific results are published and shared. These shortfalls have implications for reproducibility, because scientists who wish to reproduce research may lack the information or training they need to do so.
\end{quote}
}
Given the pioneering role of the behavioral and social sciences in the promotion of novel methodologies to improve the credibility of research, it is promising to analyze the lessons learned from this field and adjust strategies for Computer Science, AI and ML.

The authors of this position paper come from both computer science and the behavioral and social sciences, and collaborate on Replication Markets -- one part of the larger DARPA-funded program on Systematizing Confidence in Open Research and Evidence (SCORE) \citep{score}. We are one of two teams using crowdsourcing to rate the credibility of  \textbf{over 6,000} social and behavioral science research claims, by predicting the chance they would replicate successfully, in a high-power test. 

The power of human prediction has been studied in many contexts (e.g. \citep{tetlock2016superforecasting}), but less so in the replication and assessment of scientific results. If we can scalably assess the quality of a published article (or its associated claims) via replicability (or reproducibility) scores, this could help readers assess reliability when making decisions, rather than assuming that all published studies are the same. Secondly, these scores allow us to prioritize expensive replication efforts towards maximum information, minimum regret, or some other goal.. Lastly, these assessments can serve as an incentive to motivate authors to better ensure their replicability.

After discussing how forecasting can leverage the information gained from relatively labor and resource-intensive replications, we will then review opportunities and challenges of using these approaches to monitor and improve the credibility of research areas in Computer Science, AI and ML. We end with details of our efforts in the DARPA SCORE project.

\section{Related Work}
\paragraph{Replication crisis in science}
As noted above, recent works have shown relatively low replication rates for various social sciences, in both ``one-word" journals (e.g. \emph{Science} and \emph{Nature}) and more specialized ones. Despite some variation between fields, their overall low replication rates suggests unreliable practices \citep{christensen2018transparency,rahal2015estimating,baker2016there}

\paragraph{Replication in AI}

AI is no exception to concerns about replicability. Several reports claim the stated performance of a documented ML method can not be (fully) reproduced, including deep reinforcement learning \citep{henderson2018deep} and generative adversarial networks \citep{lucic2018gans}. There are other critiques in a more generic computation and machine learning context \citep{Freire:2012:CRS:2213836.2213908,gundersen2018state,pmlr-v97-bouthillier19a}.

\citep{gundersen2018state} surveyed state-of-the-art results in two premier AI venues AAAI and IJCAI. Instead of performing replication studies over a selected pool of 500 published articles, the authors identified a set of relevant proxy signals and determined a replication score from those. 

\paragraph{Prediction market for replication}

A prediction market is a market place for strategic agents to trade their private information \citep{hanson2003combinatorial,hanson1995could,wolfers2006prediction}. Prediction markets have shown to be able to aggregate information efficiently. Particularly, Hanson promoted the idea of using a prediction market to build a gambling place for ``saving science" \citep{hanson1995could}. A key property of a prediction market is to incentivize agents to truthfully report their beliefs. 

We next review several prediction markets on scientific replications.

\section{Prediction Markets for Replication Elicitation}
A prediction market allows participants to trade contracts that will eventually pay out based on the true outcome. A \textit{replication market} pays based on actual attempted replications. (Typically we require the replication to be higher power than the original study, to reduce false negatives.) Contracts are typically ``Pays \$1 if Yes'' or ``Pays \$1 if No,'' so the current price is the market's estimate of the probability for that outcome. A single replication is only a proxy for truth, but one can perform a Bayesian update \citep{dreber2015using}.


Four of the large-scale replications in the social and behavioral sciences were accompanied by forecasting studies. These studies aimed to investigate if the research community can forecast replication outcomes. In the first study \citep{dreber2015using}, we elicited forecasts on 41 of the replications conducted as part of the Replication Project Psychology (RPP; Open Science Collaboration, 2015). Studies 2 and 3 combined replications and forecasting, for 18 studies in experimental economics \citep{camerer2016evaluating}, and for 21 experimental social science studies published in Nature and Science \citep{camerer2018evaluating}. The fourth study \citep{forsell2018predicting} was conducted together with the ManyLabs 2 study \citep{klein2018many}, and elicited forecasts for the replicability of 24 classical and contemporary psychology studies.

For these 104 claims, we elicited forecasts through un-incentivized surveys followed by prediction markets. The markets ran for about 2 weeks. Participants were recruited from the relevant research communities using mailing lists and blog posts. In total, using a simple binary criterion in interpreting forecasts, prediction markets correctly predicted the replication outcomes for 73\% of 104 studies. Average survey responses were highly correlated with market forecasts but were somewhat less accurate. The forecasting projects demonstrated that prediction markets are suitable tools to elicit forecasts about the credibility of research findings from the research community.

\section{The Challenges and Possibilities of Forecasting for AI Replication}

Existing literature on assessing replication in AI has focused on tabulating proxy signals of replications for each of the published articles \citep{gundersen2018state}. While this is a very inspiring and clever way to deal with the massive amount of papers published at AI venues, it is 
\begin{itemize}
\item not a direct assessment of the papers themselves;
\item not scalable for the exponentially growing number of submissions and published articles in AI.
\end{itemize}

\noindent We next will explain why we think forecasting might be a good fit for assessing AI replicability. In particular, we think that systematic claim sampling, and crowd forecasting are relevant. 
 
\subsection{Definition of Replication}

Although the National Academies \citep{NAP25303} has recently tied "reproduction" to running the same code on the same data, and "replication" to new data and/or methods, the AI community has typically reversed them. Regardless, we need to consider how replications will be performed, what counts as success, and what is being tested. Following \citep{gundersen2018state}, we note three criteria for reproducibility in AI :
\begin{itemize}
    \item Method reproducibility 
    \item Result reproducibility
    \item Inferential reproducibility
\end{itemize}
\noindent Each of the definitions can form a forecasting question:
\begin{center}
\emph{``Is the paper/claim method/result/inferential reproducible?". }    
\end{center}
\noindent For all three criteria of replication, we argue that forecasting provides a sensible, useful and scalable way of collecting assessments. 

\emph{Method reproducibility} is the easiest to check, in AI. In light of the reproducibility crisis, major conferences in AI/ML (e.g., ICML, NeurIPS, AAAI, ICLR) encourage authors to share and publish their codes. This practice allows an easy way to check replication in AI via re-running the codes shared/published by authors. Though a relatively easy task, often people lack incentives to report the results of such attempts. Our incentive structures can help - though by itself this is not ``forecasting''. Similar to NAS' ``computational reproducibility.''

\emph{Result reproducibility} requires testing promoted methods over a different set of data. This task is suitable for our crowdsourcing approaches. With properly built incentives, we can hope to motivate relevant participants (e.g., graduate students) to test published methods over different datasets they have access to, and report their results, via either participating in our prediction market or the surveys.

\emph{Inferential reproducibility} requires replicating the main claims made in the paper with a different implementation and on a different dataset. This is the most strict definition of replication in AI and is a much trickier procedure to verify the replication of a reported result. Here in particular it would be useful to forecast widely, and check a subset. Forecasting provides a cheap and quick way to estimate a confidence score for this type of replication.

Gunderson \& Kjensmo emphasize independent teams working only from the \textit{documentation}. Their goal is to ensure results are due the AI method, not irrelevant details like hardware or environment, so they do not require a new dataset:

\begin{quote}
    Reproducibility in empirical AI research is the ability of an \textit{independent research team} to produce the same results using the same \textit{AI method} based on the \textit{documentation} made by the original research team.
\end{quote}

\noindent This also takes time, and would benefit from informed forecasts, a subset of which are checked.

\subsection{Replication Market for AI} 

\paragraph{Claim extraction}

Published articles typically present multiple major conclusions and claims, including tackling open problems, presenting computationally efficient solutions (not necessarily solving a new problem), outperforming benchmark solutions, etc. Most of the current studies require manual extraction, though SCORE appears to be building in some automation. \emph{Automated claim extraction} would greatly benefit replication studies. 

We can build one replication market for each extracted claim, asking participants 
\emph{``Is [this claim] method/result/inferential/\ldots  reproducible?". }   Typically we have used binary outcomes, but it's also possible to forecast metrics like accuracy, or multiple outcomes.  Once the markets settle, we can select a subset to test. For SCORE, this was a stratified random sample. But if we adopt a decision-market scoring rule (SCORE, Section 5.2), we can use the market results to select the claims to test, according to our goals. 

\paragraph{Meta claims} Besides forecasting replication questions for individual claims, we also find it useful to collect forecasts for \emph{meta claims} on the overall replication rate for a field. For instance, we can ask participants to forecast ``What is the average replication rate in Natural Language Processing?".

\paragraph{Survey approach} Prior to or instead of markets, we can use surveys. Surveys open up possibilities for weighted forecasts, including using peer prediction approaches (e.g., SSR, Section 5.3) to incentivize participation. Existing platforms such as openreview ({\color{blue}\url{https://openreview.net/}}) provide a very convenient way to obtain/crowdsource opinions and comments about a scientific report, but these platforms in general lack a mechanisms to properly motivate voluntary participation. 

\subsection{Implications for AI}

Forecasting might help solve the peer review shortage in major AI/ML conferences. AI conference submissions have been growing exponentially fast. For instance, AAAI 2020 received 8,800+ submissions. While this shows enormous interest in AI, it poses a substantial reviewing challenge. With an overwhelming number of papers to review, more reviewers will be needed, causing the average qualification of reviewers to decrease.
 
Quick replication forecasts can provide an informative signal to the program committee to make early ``desk rejection" decisions or perhaps to prioritize spot-checking the studies rated non-replicable. (When market results affect resolutions, decision-market scoring must be used to prevent improper incentives.) 

Besides forecasting articles' replications, our approach will also allow us to elicit forecasts on other proxy signals such as forecasted number of citations in the next few years. This information can also be elicited during pre-submission time, and inform the program committee.

\paragraph{Improve the progress of AI}

Leaving the judgment of scientific credibility in AI to the general crowd sounds dangerous. But the availability of predictions about a claim's replicability may help the community and editors better judge the value of a reported claim  \citep{DellaVigna428}. For instance, having a sense of the research community's expectation gives researchers a certain form of ``prior" for the corresponding study, which helps calibrate the surprises of a reported claim. Because forecasts may be judged, participants have some incentive to report truthfully.

\section{DARPA SCORE}
In 2019, DARPA announced the SCORE program to create ‘confidence scores’ for a large pool (3,000) of publications from the social and behavioural sciences. The motivation and goal of SCORE states as follows:

\begin{quote}
   \emph{The Department of Defense (DoD) often leverages social and behavioral science (SBS) research to design plans, guide investments, assess outcomes, and build models of human social systems and behaviors as they relate to national security challenges in the human domain. However, a number of recent empirical studies and meta-analyses have revealed that many SBS results vary dramatically in terms of their ability to be independently reproduced or replicated, which could have real-world implications for DoD’s plans, decisions, and models.} 
\end{quote}

Although SCORE includes both the replicating teams and teams developing machine learning models to score an article or claim, our team focuses on eliciting human beliefs about a variety of metrics relevant to replication. For instance, for each field of studies, we are interested in knowing people’s estimates of the average replication rate.

For each particular article, another team (the Center for Open Science) first extracts the main claims. We then elicit people's following two estimates: 
\begin{itemize}
\item Estimate of its probability of being replicated, defined as a statistically significant result in the same direction, when using substantially the same method and analysis, but new data and likely a much larger sample size (direct replication). 
\item Estimate of its probability of being replicated if we did exactly the same analysis on a novel but ``found" dataset, such as GDP figures from later years, or a different country (data replication).
\end{itemize}

A challenge in this project is that only a small fraction of these studies will actually be selected for replication  ($\sim 100$ direct and $\sim 150$ data replications), thus forecasting approaches to create these confidence scores had to function with ground truth becoming available only for a small fraction of the forecasts. We therefore used two different approaches to deal with the high rate of unverifiable outcomes, decision markets \citep{chen2011decision}, and surrogate scoring rules (SSR) \citep{liu2018surrogate}.

\begin{figure}[!ht]
\begin{center}
\includegraphics[width=0.45\textwidth]{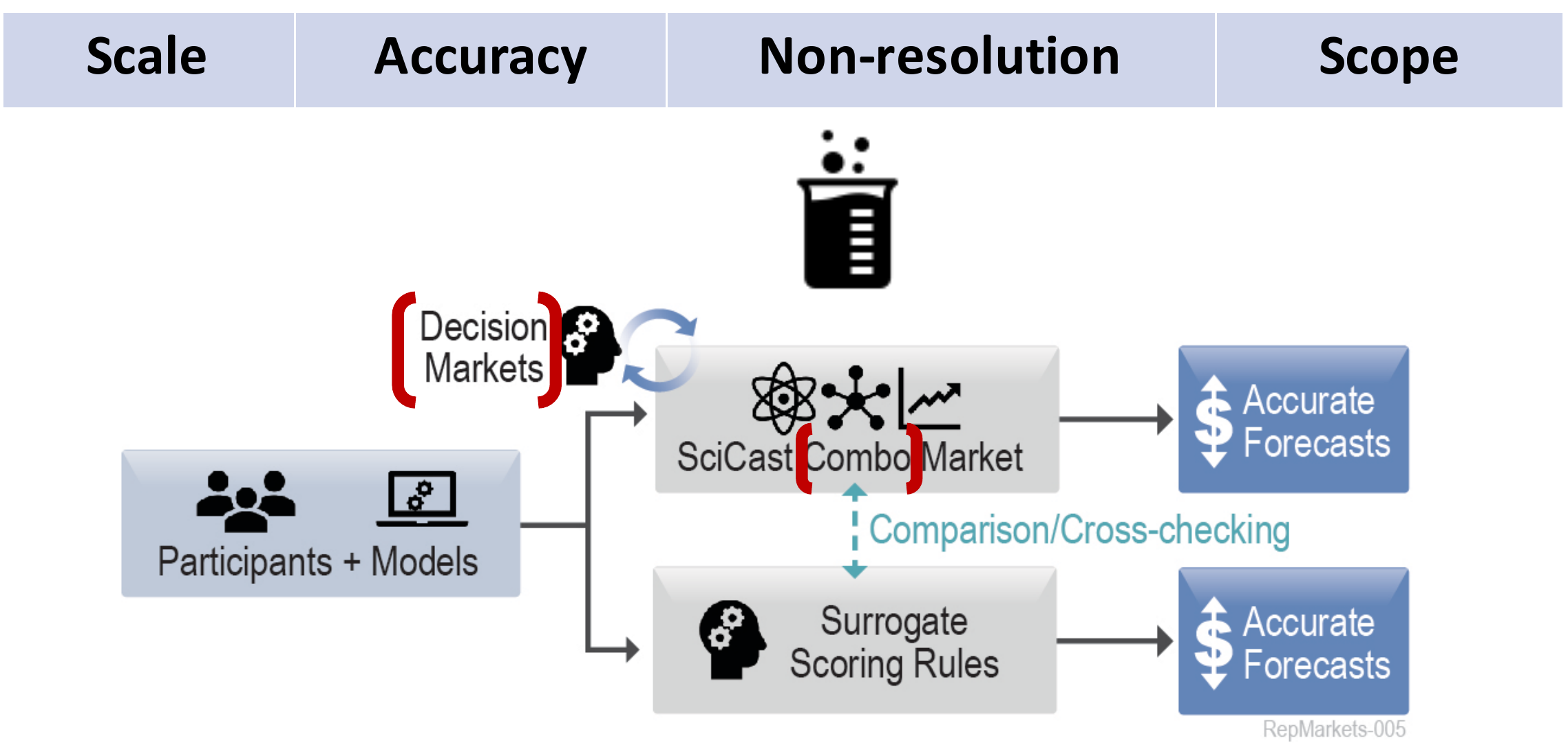}
\caption{Overview of our project for SCORE, highlighting the three components and challenges.}\label{HFC}
\end{center}
\end{figure}

Our project features 
\begin{itemize}
\item a continuation of using prediction market for eliciting estimates of replications;
\item a decision market that uses crowd forecast to inform targeted replication efforts; 
\item a peer prediction approach that incentivizes participation without using ground truth outcome of the replicability of articles.
\end{itemize}

\subsection{Replication Market}

\begin{figure}[!ht]
\begin{center}
\includegraphics[width=0.45\textwidth]{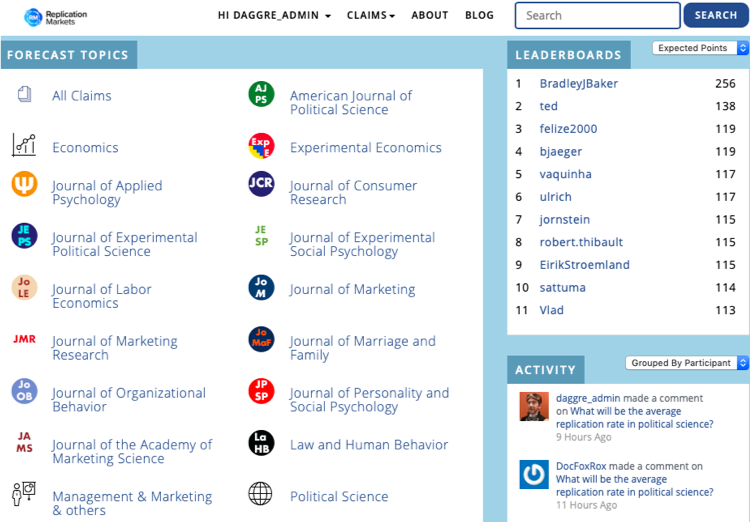}
\caption{Overview of replication market we built.\\ {\color{blue}\url{https://www.replicationmarkets.com/}}}\label{RM:interfact}
\end{center}
\end{figure}

A primary component of our project is a prediction market, which we call a replication market. Fig. \ref{RM:interfact} shows the welcome page of the market we built. Once they arrive, participants can choose the topics and their interests and find the relevant markets/claims to engage. Our front page presents a leaderboard that shows the top performers. 

For our first round of replication markets, we successfully recruited 508 eligible participants (finished registration surveys), and had in total 218 active traders, who made 1,729 trades. We now have over 1,500 eligible participants, and about 150 active.  We completed surveys and markets for both Round 1 (~300 claims) and Round 2 (~200 claims).  

\begin{figure}[!ht]
\begin{center}
\includegraphics[width=0.2\textwidth]{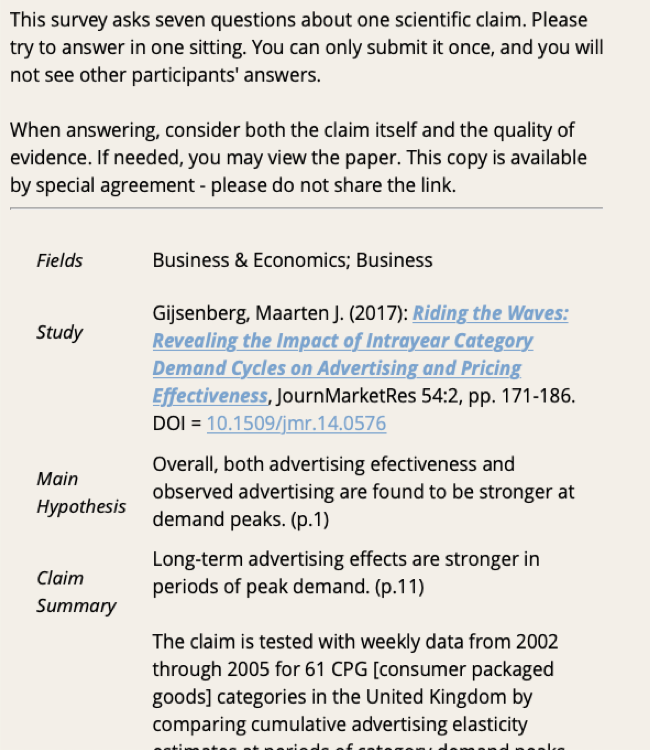}
\includegraphics[width=0.2\textwidth]{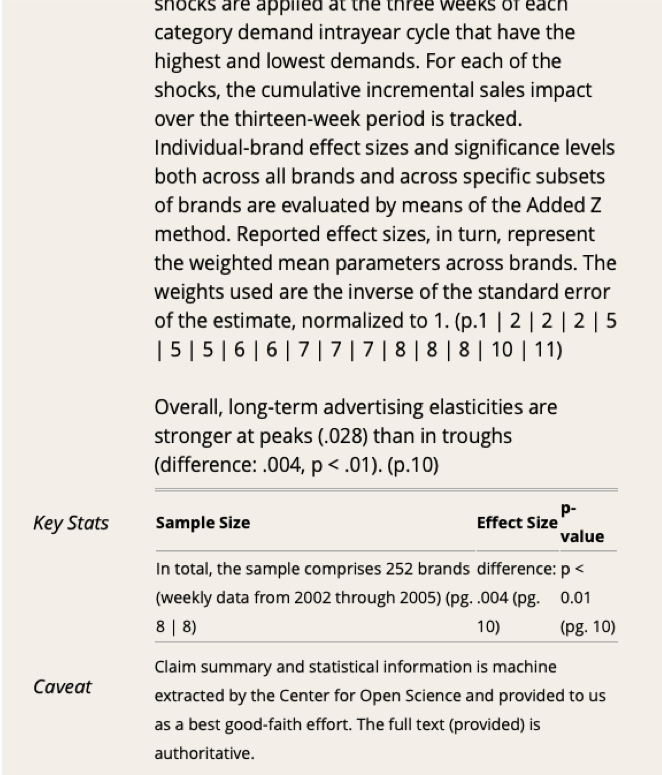}
\caption{Claims, key statistics and pointers presented to our participants.}\label{RM:claim}
\end{center}
\end{figure}

Claims and the associated key statistics were extracted to present to our participants to help with their assessments. In each Round, the first week is individual surveys, followed by two weeks where all traders can see and select all claims.  See Fig. \ref{RM:claim} for an example.
Both the survey and the market present detailed claims summaries, with the option to click through to the full text.

\subsection{Decision Market for Targeted Spot-check}

In a prediction market, participants are rewarded according to the closeness of their predictions to the future resolved event outcome. They are hence incentivized to provide their most accurate assessments on the event. With limited resources, we simply won't be able to check every paper's replication. In other words, our questions are not going to be resolved, and the majority of the prediction markets won’t pay off. This creates a challenge for using prediction markets to obtain an accurate assessment for all studies, including the non-resolved ones: market participants, conjecturing or knowing which studies will be selected may focus their effort on predicting these studies because their bets in other markets likely won’t materialize.

Moreover, we hope that in the future, accurate confidence scores will inform replication decisions, to more efficiently allocate scarce replication resources over a large number of studies (e.g. attempt replication for more studies whose confidence scores are low). However, when market predictions may influence which studies will be selected for replication, a market participant may manipulate market predictions so that studies for those markets where he has higher potential payoffs are attempted for replication. The interaction between predictions and decisions can affect the accuracy of the predictions. 

We use decision markets to address both challenges. A decision market is a set of prediction markets coupled with a stochastic decision rule. For each study, there is a prediction market eliciting forecasts on the probability that the result of the study will be reproduced if a replication of the study is attempted. The decision rule specifies the probability to conduct a replication for each study – a simple rule would be a flat 5\% chance. 

To ensure incentive compatibility, if a replication for study $i$ is attempted, the decision market scales market participants' payoffs in the prediction market for study $i$ by the inverse of the probability that the study is selected for replication, i.e. by $1/p_i$. Markets corresponding to studies that are not selected do not pay off. When every study has a non-zero probability of selection, market participants are incentivized to maximize their expected payoff in each market as if there will be replication attempts on all studies. Such a decision market preserves incentive compatibility.

\subsection{Survey method: a Peer Prediction Approach}
The market approach has two limitations for our SCORE project:
\begin{itemize}
\item The markets will need the ground truth outcome of each article’s reproducibility to close. 
\item Even for the articles chosen for reproduction, the outcomes only arrive after a year. The delay in paying out will potentially hurt participants' engagements. 
\end{itemize}

To improve the incentives and participation, we implemented and tested a novel surrogate scoring technique \citep{liu2018surrogate,wang2019forecast} designed to yield incentive-compatible estimates on non-resolving questions. Eliciting confidence scores for scientific claims is a problem of information elicitation without verification. Replications can better approximate the ground truth, but only a small fraction of studies can be attempted for replications. Hence we test our surrogate scoring technique. 

\subsubsection{Surrogate scoring rule}

When we know the ground truth, we can use a strictly proper scoring rule (e.g. the Brier score) incentivizes people to offer their most accurate predictions. But when the ground truth is not available, hope is not lost. If there is a surrogate for the ground truth, with known error rates, then surrogate scoring rules can score predictions using this proxy truth to provide the same incentive as using a proper scoring rule with access to the ground truth. Suppose we are predicting whether a particular scientific claim is true (1) or false (0). Let $o$ for “outcome” be the unobservable ground truth. If we will have access to surrogate ground truth $os$ where we know the error rates, $e_0=P(os=1|o=0)$ and $e_1= P(os=0|o=1)$, then we can define the following surrogate scoring rule to score a prediction $p$:
$$
SSR(p,os) = \frac{(1-e_{1-os}) \cdot S(p,os) - e_{os} \cdot S (p, 1-os)}{1-e_{0}-e_{1}}
$$

A nice property of the above surrogate scoring rule is that when a prediction p is evaluated against the surrogate outcome os, the surrogate score in expectation equals the Brier score for the prediction, evaluated against the true outcome o. Thus, predictions are as if scored against the ground truth using the Brier score. 

The remaining challenge is how to find a surrogate ground truth with known error rates. Our SSR method achieves this under some assumptions by estimating a surrogate ground truth and its error rates from the predictions of all participants across multiple questions. The error of the estimation converges to 0 as the number of predictions increases. Hence, the average surrogate score that a participant gets converges to the average Brier score that he would get if ground truth were available. Fig. \ref{HFC} shows how surrogate scores (no access to ground truth) compare with Brier scores (with access to ground truth) on an experimental data set on predicting whether prices of some art pieces exceed a certain value. 

\begin{figure}[!ht]
\begin{center}
\includegraphics[width=0.45\textwidth]{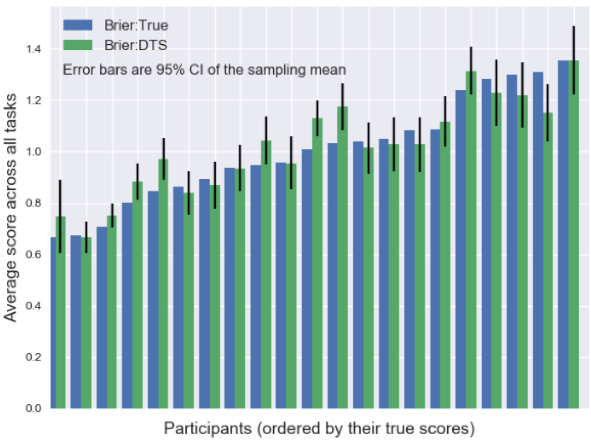}
\caption{Surrogate scores (denoted as DTS) v.s. true scores of participants' performance}\label{HFC}
\end{center}
\end{figure}

Because surrogate scores recover Brier scores in expectation and Brier scores quantify the quality of predictions (i.e. more accurate predictions lead to better Brier scores in expectation), we also want to explore using surrogate scoring techniques to post-process elicited predictions to obtain more accurate CS. In forming an aggregated CS, predictions with higher surrogate score will be weighted higher, while the exact weights depend on the estimated error rates of the surrogate ground truth. 

\begin{figure}[!ht]
\begin{center}
\includegraphics[width=0.45\textwidth]{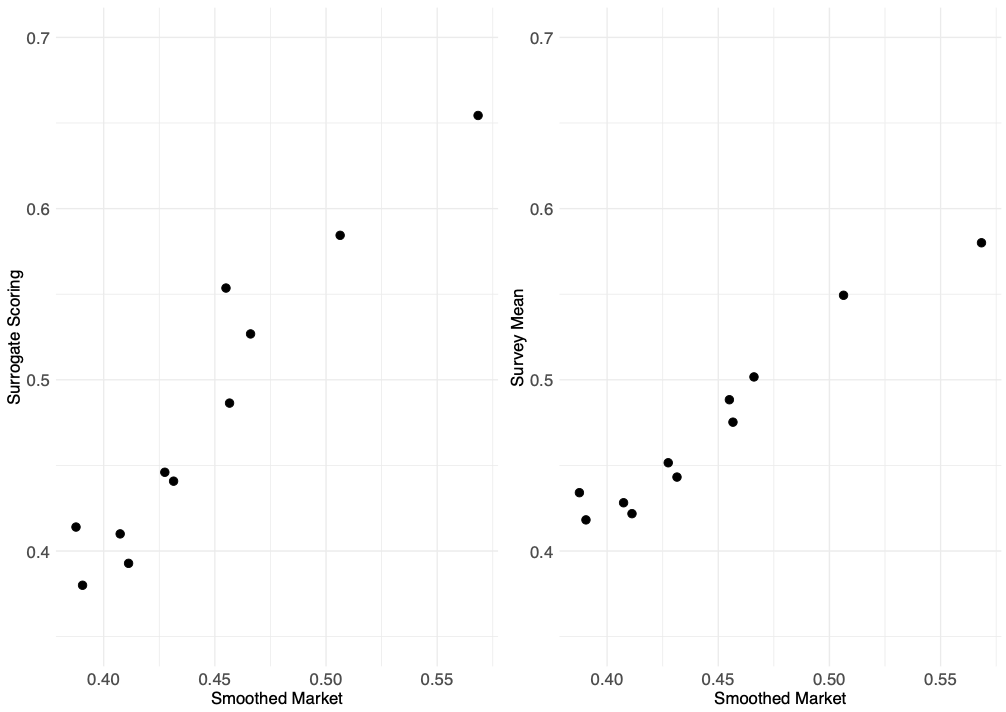}
\caption{A cross validation check of SSR elicited and aggregated survey answers and market responses. }\label{SSR:score}
\end{center}
\end{figure}

Due to the lack of ground truth verification, we cannot claim too much about whether we have identified the real best experts. But we performed a cross validation check between the SSR weighted aggregation with the market aggregated outcomes. We generally observe a strong correlation among these responses (Fig.\ref{SSR:score}). This is a positive signal about SSR's informativeness, if we believe both surveys and markets are correlating with the true performances in a certain way.

\subsection{Takeaways}

Through our previous studies, our main takeaways are 
\begin{itemize}
\item To fully embrace the wisdom of crowd, we need to use approaches that scale well and automatic – i.e. market approaches ( prediction and decision markets) and SSR. 
\item Better engage participants. With actively recruiting and engaging participants, we are able to maintain a relatively diverse and informative population to contribute forecasts. Our efforts in engaging participants include sending monthly prizes for winners in our survey studies (highlighted later), maintaining leaderboard of top performers, sending monthly emails to remind participants. 
\item Maintaining a diverse population also means that we have a participation population with different backgrounds. We design simple and easy-to-use interface, tutorials, as well as FAQ pages to let our participants understand the basic concepts of our project and the definitions of questions we are eliciting forecasts for. 
\end{itemize}

\subsubsection{Monthly prize giving}

Besides active outreach activities, we gave out prizes for participation monthly. This is a new component we tested in comparing to other forecasting efforts. This is a challenging task to do as we do not have ground truth outcomes to select the winner in forecasting until much later in the program. Waiting until then might significantly discourage the participation. The design of SSR (Section 4.3) allows us to do so. 

Specifically, our questions arrive in batches, and each month (we also call this each round) has 30 batches of questions for users to answer, with each batch consisting of claims to forecast on. SSR scores are computed for each batch of questions every month for every participant. We then use SSR to rank participants and give prizes to the top performers of each batch. We do observe a boost in participation and engagement after we gave out prizes. This is further evidenced by tweets from our participants mentioning that they won awards, acknowledging winning prizes on personal websites and blogs. We have summarized the statistics of our given prizes in the first four rounds in the following article: {\color{blue}\url{https://www.replicationmarkets.com/index.php/2020/01/02/robustness-of-monthly-prizes/}}




\section*{Acknowledgement}
This work is funded by
the Defense Advanced Research Projects Agency (DARPA) and Space and Naval Warfare Systems Center Pacific (SSC Pacific) under Contract No. N66001-19-C-4014. The views and conclusions contained herein
are those of the authors and should not be interpreted as necessarily representing the official policies, either
expressed or implied, of DARPA, SSC Pacific or the U.S. Government. The U.S. Government is authorized to reproduce and distribute reprints for governmental purposes notwithstanding any
copyright annotation therein.

\bibliographystyle{aaai}
\bibliography{mybib}

\end{document}